\documentstyle[aps,twocolumn,floats]{revtex}
\begin{document}

\tighten
\draft
\twocolumn[\hsize\textwidth\columnwidth\hsize\csname
@twocolumnfalse\endcsname

\title{Back Reaction Of Perturbations In Two Scalar Field Inflationary Models}

\author{Ghazal Geshnizjani$^{1)}$ and Robert Brandenberger$^{2)}$}

\address{1) Department of Physics, Brown University, Providence, RI 02912,
USA\\
E-mail: ghazal@het.brown.edu}

\address{2)Department of Physics, Brown University, Providence, RI 02912, USA\\
E-mail: rhb@het.brown.edu}

\maketitle

\begin{abstract}

We calculate the back-reaction of long wavelength cosmological
perturbations on a general relativistic measure of the local
expansion rate of the Universe. Specifically, we consider a
cosmological model in which matter is described by two scalar
matter fields, one being the inflaton and the other one
representing a matter field which is used as a clock. We analyze
back-reaction in a phase of inflaton-driven slow-roll inflation,
and find that the leading infrared back-reaction terms
contributing to the evolution of the expansion rate do not vanish
when measured at a fixed value of the clock field. We also analyze
the back-reaction of entropy modes in a specific cosmological
model with negative square mass for the entropy field and find
that back-reaction can become significant. Our work provides
evidence that, in general, the back-reaction of infrared
fluctuations could be locally observable.

\end{abstract}

\vspace*{1cm} ]

\section{Introduction}

Because of the nonlinear nature of the Einstein equations,
the ansatz for metric and matter used all the time in early
Universe cosmology, namely a homogeneous and isotropic
background space-time plus fluctuations which average to
zero on the spatial hypersurfaces (set by the background
cosmology) and which obey the linear fluctuation equations, does
not obey the Einstein equations at second order in the
expansion parameter (the fractional amplitude of the linear
fluctuations). Terms which have to be added to the metric
such that the Einstein equations are satisfied to second
order are called {\it back-reaction terms}.

The back-reaction of short wavelength gravitational waves on
an expanding Friedmann-Robertson-Walker (FRW) cosmology
is a well-established subject \cite{BHI}. We insert the
ansatz for the metric consisting of linear fluctuations
$\delta g_{\mu \nu}$ about the FRW background $g^{0}_{\mu \nu}$
\begin{equation}
g_{\mu \nu} = g^{0}_{\mu \nu} + \delta g_{\mu \nu}
\end{equation}
into the Einstein equations
\begin{equation}
G_{\mu \nu} = T_{\mu \nu} \, ,
\end{equation}
$G_{\mu \nu}$ and $T_{\mu \nu}$ denoting the Einstein tensor
and the energy-momentum tensor, respectively,
expand to second order, use the fact that the linear
fluctuation equations are satisfied to cancel the linear
terms, and move all quadratic terms to the right hand
side of the equation. These quadratic terms are
interpreted as an effective energy-momentum tensor $\tau_{\mu \nu}$
(``pseudotensor'') for fluctuations. The resulting
equation
\begin{equation} \label{breq}
G_{\mu \nu}(g^{br}) = T_{\mu \nu}(\varphi^{br}) + \tau_{\mu \nu}
\end{equation}
(up to terms of third and higher order)
can be interpreted in the sense that the
fluctuations carry energy and pressure and hence effect
the background space-time in which they are defined. Note, that in
order for this equation to be satisfied, we need to add quadratic
correction terms (``back-reaction terms'') to the metric and
the matter (which explains the superscripts in the above equation).
In particular, the homogeneous component of the metric obtains
a correction term of quadratic order which can be extracted from
(\ref{breq}) by taking spatial averages (see \cite{ABM,Abramo} for
more details).

Since, in the context of inflationary and
post-inflationary cosmology, the scalar metric fluctuations
(fluctuations coupled to energy density and pressure perturbations)
are believed to dominate over the effects of gravitational waves, it is
of great interest to study the back-reaction of these cosmological
perturbations. Furthermore, in inflationary cosmology the phase space
of infrared modes (defined as modes with wavelength greater than the
Hubble radius) is growing over time, whereas the phase space of
ultraviolet modes does not grow, and since the amplitude of the
associated metric fluctuations of these infrared modes does {\it not}
decrease in time (see e.g. \cite{RA} for a comprehensive overview
of the theory of cosmological fluctuations and \cite{B03} for a recent
introductory overview), the back-reaction of these infrared modes
may be very important.

Rather recently, the effective energy-momentum tensor
formalism of \cite{BHI} was generalized to study the
back-reaction of infrared modes on the spatially homogeneous component
of the metric \cite{MAB,ABM}. The surprising result
which was found is that in a slow-roll inflationary
background, the infrared contribution to $\tau_{\mu \nu}$
has the form of a negative cosmological constant, with the absolute
value of the energy density increasing in time as the phase space
of infrared modes grows and the amplitude of the gravitational
potential remains large (in fact the amplitude is slowly increasing).
A similar result was found in an earlier study by Tsamis and
Woodard \cite{WT}, in which the back-reaction of long wavelength
gravitational waves in a de Sitter background was studied
\footnote{The back-reaction of small-scale (i.e. smaller than
the Hubble radius) cosmological perturbations has been
considered in
\cite{Futamase1,Futamase2,BF,Futamase3,SH,Ruus,Jelle,Buchert},
and in  \cite{BE,TF} in the context of Newtonian cosmological
perturbation theory. Note also that
Nambu \cite{Nambu1,Nambu2,Nambu3}
is developing a program to compute back-reaction effects on the
spatially averaged metric using the renormalization group method.}. The
results of \cite{MAB,ABM} were confirmed in \cite{aw0} using
very different techniques. On the basis of these calculations, it was
argued \cite{MAB,ABM,aw0} that gravitational back-reaction in
scalar field-driven inflationary models, calculated up to quadratic order
in perturbations and to leading
order in the long wavelength expansion and in the slow roll approximation could
decrease the expansion rate of the universe
and potentially solve the cosmological constant problem \cite{bran4,bran5} (see
also \cite{WT,atw} for similar suggestions in the context of back-reaction
studies of gravitational waves, and \cite{Mottola} for earlier ideas
concerning the instability of de Sitter space).

However, as was emphasized by Unruh \cite{unruh}
\footnote{We are also grateful
to Alan Guth, Andrei Linde and Alex Vilenkin for detailed private
discussions on these points.}, the approach of
\cite{MAB,ABM,WT} has several shortcomings. First of all,
due to the nonlinear nature of the Einstein equations, calculating
an ``observable'' from the spatially averaged metric will not in
general give the same result as calculating the spatially averaged value
of the observable. More importantly, the spatially averaged metric is
not a local physical observable. An improved analysis of
gravitational back-reaction on the locally measurable expansion
rate of the Universe starts with
identifying a local physical variable which describes this
expansion rate, then calculating the back-reaction
of cosmological perturbations on this quantity, and only at the end
taking an expectation value of the result. It is important
to fix the hypersurface of averaging by a clear physical
prescription in order to remove the possibility of being
misled by coordinate artifacts which are not physically
measurable. Such an approach was first suggested by Abramo and
Woodard in \cite{aw2}.

In a first paper \cite{GB1}, we considered a simple variable describing
the local expansion rate and calculated this variable to second
order in the metric fluctuations in a model with a single matter
field, the inflaton field $\varphi$. We then evaluated the
leading contributions of infrared fluctuation modes to this
variable. When evaluated at a fixed background time $t$, we
found that the back-reaction effects were present. However, when
evaluated at a fixed value of the matter field $\varphi$, the only
physical clock available in this simple system, the dependence
of the expansion rate on the clock time took on exactly the same
form as in an unperturbed background. Thus, the leading infrared
back-reaction terms had no locally measurable effect in this
system. Similar conclusions were reached in \cite{aw3}.

However, the fact that the infrared back-reaction terms did not
vanish when evaluated at a fixed background time leads us to
the conjecture that they will have a locally measurable effect
provided we have a clock different from the inflaton field
$\varphi$ itself \footnote{To explain the effects of back-reaction
in heuristic language one needs to introduce a second matter field:
the back-reaction effect is the fact that the expansion rate is
different in the presence of fluctuations compared to the result
in the absence of fluctuations {\it when measured at the same
value of the temperature of the CMB}.}. The simplest way to
introduce an independent physical clock is to add a second scalar
matter field $\chi$ to the system. This field $\chi$ represents
our observed matter fields.

In this paper, we evaluate our variable describing the local
expansion rate of the Universe to second order in the
fluctuations in the slow-rolling phase of an inflationary
Universe with two matter field, the inflaton $\varphi$ which dominates
the energy-momentum tensor and a spectator (clock) field
$\chi$. We find that, in general, the leading infrared
terms {\bf do not} vanish when evaluated at a fixed value of
the clock field $\chi$. Thus, in this system infrared back-reaction
is for real \footnote{Another system in which infrared back-reaction
is for real was discussed in \cite{aw4}.}. We also discuss some
subteties concerning this conclusion. The physical effects
of this infrared back-reaction remain to be investigated \cite{GB3}.

The outline of this article is as follows:

In the following section we review the construction of the
variable $\Theta$ which we propose to use to describe the local
cosmological expansion rate, and we summarize the results of our
previous paper \cite{GB1} in which this variable was computed in a
cosmological model with a single scalar matter field.

 In Section 3, we derive
the expression for $\Theta$ in a model with two matter fields in
terms of the values of the metric fluctuations (more specifically,
in terms of the gravitational potential $\Phi$ which characterizes
the fluctuations in the longitudinal gauge).

In Section 4, we derive an expression for the change of $\Theta$
over a short time during which the clock field $\chi$ changes by a
small amount, in terms of physically measurable quantities. we
investigate the behavior of back-reaction in scenario where
adiabatic modes are dominant and $\chi$ is a rapidly oscillating
scalar filed where the local energy density of $\chi$ is our
clock.
We show that when evaluated during the slow-roll phase of an
inflationary Universe, there are non-vanishing back-reaction
contributions from infrared modes.

Then, in section 5, we consider a two matter field model motivated
by hybrid inflation, where it is expected that the entropy
fluctuations are large. We calculate the back-reaction effect in
terms of proper time, as a clock and we show that back-reaction of
entropy modes could be significant.

 In the
final section we summarize and discuss our results and point out
issues which remain to be resolved.

\section{Local Back-Reaction in Single Field Models}

We begin by reviewing the construction of the variable $\Theta$
measuring the local cosmological expansion rate: We consider
the velocity four-vector field $u^{\alpha}$ tangential to
a family of world lines of a matter fluid in a general space-time.
This four-vector is normalized such that
\begin{equation} \label{norm}
u^{\alpha}u_{\alpha} \, = \, 1 \, ,
\end{equation}
where $\alpha$ runs over the space-time indices. In terms of
this four-vector, the local expansion rate $\Theta$ is
defined by
\begin{equation}
\Theta \, \equiv \, u^{\alpha}_{\,;\alpha} \, .
\end{equation}
It represents the local expansion rate of the tangential surfaces
orthogonal to the fluid flow.

For a homogeneous Universe
with scale factor $a(t)$, the Hubble expansion rate $H$ is related
to $\Theta$ via
\begin{equation} \label{Hubble}
H \, \equiv \, \partial_{t}a/a \, = \, \frac{1}{3}\Theta \, .
\end{equation}
Thus, for a Universe with fluctuations, we use
$\partial_{t}S/S \equiv \frac{1}{3}\Theta$
to define the ``locally measured'' Hubble expansion rate.
This variable is a mathematically simpler object than the
one recently introduced in \cite{aw2}, which involves the
integral along the past light cone of the observation point of a
function which itself is rather difficult to calculate.

In \cite{GB1}, we considered a theory with a single matter
field $\varphi$ and evaluated $\Theta$ up to second order in
the metric fluctuation variable $\Phi$ (defined later in
(\ref{metric})). When expressed as a function of physical
time $t$, and dropping all spatial gradient terms, the
result obtained is
\begin{equation} \label{form2}
\Theta \, = \, 3\frac{\dot{a}}{a}(1-\Phi+\frac{3}{2}\Phi^{2})
-3(\dot{\Phi}+\dot{\Phi}\Phi)
\end{equation}
Thus, at first glance it looks like the local expansion rate
is different in the presence ($\Theta \neq 0$)
than in the absence ($\Theta = 0$) of fluctuations.
However, the above conclusion is premature since the background
time $t$ is not a physical quantity.

In order to draw conclusions about the local measurability of
the back-reaction of infrared fluctuation modes, it is necessary
to express the final result in terms of a locally measurable
quantity, a clock. The only such clock available in the single
matter field model considered in \cite{GB1} is the inflaton
field itself. Thus, we must express the metric fluctuations $\Phi$
in terms of the matter fields. This can be done by making use
of the Einstein constraint equations. Neglecting spatial
derivatives in these equations, one can show (see also \cite{niayesh})
that, also setting ${\dot \Phi} = 0$ (which is approximately the case
for each long wavelength fluctuation mode in a slow-roll inflationary
model \cite{RA}), the final result for the local expansion rate becomes
\begin{equation} \label{final}
\Theta \, = \, \sqrt{3} \sqrt{V(\varphi)}
\end{equation}
which as a function of $\varphi$ is the same as the relation for
an unperturbed background. Thus, in the single matter field model
the leading back-reaction terms from infrared modes are not
locally measurable, in agreement with the conjectures on
\cite{unruh} \footnote{Note, however, that there still could be
back-reaction of infrared terms which can change the coarse of
evolution of $\varphi$, due to the fact that new modes are
continuously exiting the Hubble radius. These effects are
calculated in \cite{AG}.}

In the presence of a second matter field $\chi$, it is possible
to use this field as a clock. In fact, using this field to
model the temperature of the CMB, it is physically more sensible
to use $\chi$ as the clock rather than the inflaton field itself.
The fact that in the single field matter model the dominant
infrared back-reaction terms {\it do not} vanish when evaluated
at a fixed background time $t$ leads us to conjecture that they
{\it will not} vanish when evaluated at a fixed value of the
new clock field $\chi$ in the two matter field model.

If we are able to verify the above conjecture, we would have
established a close analogy (already suggested in \cite{GB1})
with the analysis of the parametric
amplification of super-Hubble-scale cosmological fluctuations
during inflationary reheating. From the point of view of the
background space-time coordinates, it appears \cite{Kaiser} that
the parametric amplification of matter fluctuations on
super-Hubble scales in an unperturbed cosmological background (see
e.g. \cite{tb,kls} for a discussion of parametric resonance during
reheating) would imply the parametric amplification of the
cosmological fluctuations on these scales. However, it can be
shown that in single field models physical observables measuring
the amplitude of cosmological fluctuations do not feel any
resonance \cite{fb1,parry,lin,niayesh}. In contrast, in two field models
of inflation there is \cite{bv,fb2} parametric amplification of
super-Hubble-scale cosmological fluctuations. In this case, there is a
fluctuation mode corresponding to entropy fluctuations which
cannot locally be gauged away. This mode is (in certain theories)
parametrically amplified during reheating, and in turn drives the
parametric resonance of the super-Hubble scale curvature
fluctuations.

\section{Deriving the Expansion Rate for Models with Two Matter Fields}

The procedure for calculating $\Theta$ will be as follows:
First, we determine the velocity four-vector field
$u^{\alpha}$ for the two field model. Then,
we use the Einstein equations to express $u^{\alpha}$
in terms of the metric perturbations. Taking the
relative amplitude of the metric fluctuations as the expansion
parameter, we then calculate
$\Theta$, our local measure of the Hubble expansion rate, to second order.
After evaluating the result on a physically determined hypersurface
we can then study the back-reaction of cosmological fluctuations
on the locally measured Hubble expansion rate.
We will focus on the leading infrared contributions to back-reaction,
the terms found to dominate the back-reaction effects in
\cite{ABM,MAB,aw0}.

We consider a model with two scalar matter fields,
the inflaton $\varphi$ and the clock field $\chi$:
\begin{eqnarray}
{\cal L}
&=&\frac{1}{2}\partial^{\mu}\varphi\partial_{\mu}\varphi-V(\varphi,\chi)\\
{\cal L}^{\prime}&=&\frac{1}{2}
\partial^{\mu}\chi\partial_{\mu}\chi \, .
\end{eqnarray}
We describe the inflaton as a perfect fluid with energy density
$\rho$ and pressure $P$, considering
\begin{equation}
(\rho+P)u_{\mu}u_{\nu} - P g_{\mu\nu} \, = \,
\partial_{\mu}\varphi\partial_{\nu}\varphi-{\cal L} \, .
g_{\mu\nu}
\end{equation}
which can be justified by taking
\begin{eqnarray}
P \, &=& \, {\cal L} \,\,\,  {\rm and} \\
\rho \, &=&
\frac{1}{2}\partial_{\mu}\varphi\partial_{\nu}\varphi+V(\varphi,\chi)
\, ,
\end{eqnarray}
which leads to
\begin{eqnarray}
u_{\mu}&=&A\partial_{\mu}\varphi \, \\
A&=&\bigl(\partial^{\nu}\varphi\partial_{\nu}\varphi\bigr)^{-1/2}
\, .
\end{eqnarray}

Starting from the expression for the metric to linear order in the
fluctuations $\Phi$ (see \cite{RA} for a detailed review), we
determine the velocity four-vector field $u^{\alpha}$ to second
order, the order required to analyze the leading infrared terms in
the back-reaction to quadratic order. In linear perturbation
theory, and in the case of simple forms of matter for which there
is to linear order no anisotropic stress, the metric (in
longitudinal gauge) can be written as
\begin{eqnarray} \label{metric}
ds^{2}&=& a(\eta)^2\bigl((1+2\Phi)d\eta^{2} -
(1-2\Psi)\gamma _{ij}dx^{i}dx^{j}\bigr),\\
\gamma_{ij}&=&\delta_{ij}[1+\frac{1}{4}{\cal
K}(x^{2}+y^{2}+z^{2})]
\end{eqnarray}
where ${\cal K}=0,1,-1$  depending on whether the
three-dimensional space corresponding to the hypersurface $t =
{\rm const.}$ is flat, closed or open. In this paper we will take
it to be zero in order to simplify the calculations. The time
variable $\eta$ appearing in (\ref{metric}) is conformal time and
is related to the coordinate time $t$ via $d\eta=a^{-1}dt$. For
the forms of matter considered here, $\Psi=\Phi$ at linear order
\footnote{Even if we did not make the assumption $\Psi = \Phi$, it
turns out that at second order all infrared terms depending on
$\Psi - \Phi$ will drop out.}. The field $\Phi$ is then called the
relativistic gravitational potential.

In order to obtain the complete result for gravitational
back-reaction, we consider the Einstein equations for a perfect
fluid and additional scalar field $\chi$,
\begin{equation}
G_{\mu\nu}
=(P+\rho)u_{\mu}u_{\nu}-Pg_{\mu\nu}+\partial_{\mu}\chi\partial_{\nu}\chi-{\cal
L}^{\prime}g_{\mu\nu}
\end{equation}
(in units in which $8 \pi G = 1$), which, since
$G^{\mu}_{\mu}=-R$, will yield
\begin{equation}
-R=\rho-3P-\partial^{\mu}\chi\partial_{\mu}\chi \, .
\end{equation}
We can define a new tensor called $G^{\prime}_{\mu\nu}$ and a new
scalar $R^{\prime}$ by,
\begin{eqnarray}
G^{\prime}_{\mu\nu}&=&G_{\mu\nu}-\partial_{\mu}\chi\partial_{\nu}\chi+{\cal
L}^{\prime}\cdot g_{\mu\nu} \\
R^{\prime}&=&R-\partial^{\mu}\chi\partial_{\mu}\chi \, .
\end{eqnarray}
Then, the equations become similar to the case of a single scalar field (see
\cite{GB1}), and $\rho$ satisfies the equation,
\begin{equation}
\rho=u^{\mu}G^{\prime}_{\mu\nu}u^{\nu} \,
\end{equation}
which leads to an equation that can be solved perturbatively
for $u_{i}$ and $u_{0}$ in parallel with equation (\ref{norm})
to any desired order:
\begin{equation} \label{einstein}
G^{\prime
0}_{i}=\frac{4}{3}u^{\mu}G^{\prime}_{\mu\nu}u^{\nu}u^{0}u_{i}+
R^{\prime}u^{0}u_{i} \, .
\end{equation}

For an unperturbed Robertson-Walker metric, the four-velocity
field $u$ in comoving coordinates would be
\begin{equation}
u^{\mu}=(1,0,0,0) \, ,
\end{equation}
which can be substituted in equation(\ref{einstein}) as zeroth
order solution in an expansion in powers of $\Phi$ to calculate $u^{i}$ up to
second order,
\begin{equation}\label{ui}
u^{i^{(2)}}=\frac{G^{0i}-\partial^{0}\chi\partial^{i}\chi}{\frac{4}{3}u^{0^{3}}G^{\prime}_{00}+\frac{1}{3}R^{\prime}u^{0}}
\, ,
\end{equation}
where $G^{0i}$ is proportional to $\partial^{i}\Phi$. Using
equation (\ref{norm}) we can derive the expression for the time
component of $u^{\alpha}$ in terms of $\Phi$:
\begin{equation}
u^{0}(\eta)=a^{-1}(1-\Phi+\frac{3}{2}\Phi^{2})+\frac{1}{2}au^{i}u^{i} \, .
\end{equation}

Now that we have all components of $u^{\alpha}$, we need to take
the covariant derivative of this vector and retain all $\Phi$ dependence up
to second order. During inflation, fluctuations which are
generated on sub-Hubble scales early on during the inflationary
phase are red-shifted to scales much larger than the Hubble
radius. Thus, in this context it is of great interest to consider
the back-reaction of infrared modes. To investigate this effect we
neglect any spatial derivative term contributing to the local
expansion rate $\Theta$. Since
\begin{equation}
u^{\alpha};_{\alpha} \, = \,
\frac{1}{\sqrt{g}}\partial_{\mu}(\sqrt{g}u^{\mu}) \, ,
\end{equation}
and, according to equation(\ref{ui}), spatial components of $u$ are
proportional to spatial derivatives of $\Phi$ and $\chi$, we conclude
that, neglecting the spatial derivative terms, the expression for
$\Theta$ reduces to
\begin{equation}
\Theta \, = \, \frac{1}{\sqrt{g}}\partial_{0}(\sqrt{g}u^{0}) \, .
\end{equation}
A straightforward calculation yields the result
\begin{equation} \label{expform}
\Theta \, = \, 3\frac{\dot{a}}{a}(1-\Phi+\frac{3}{2}\Phi^{2})
-3(\dot{\Phi}+ \dot{\Phi}\Phi) \, .
\end{equation}
Where a dot denotes differentiation with respect to the coordinate
time $t$. Since $H=\dot{a}/a$, we can immediately read off the
extra terms contributing to the local expansion rate which result
from the presence of cosmological fluctuations. Hence, it follows
that if evaluated at a fixed coordinate time, infrared modes on
average lead to corrections in the expansion rate compared to what
would be obtained at the same coordinate time in the absence of
metric fluctuations. Whether this is a physically measurable
effect will be discussed in more depth in the following section.

\section{Expansion Rate in Terms of a Scalar Field as Observable }

The expression (\ref{expform}) gives $\Theta$ in terms of the
metric fluctuations. However, to determine whether this local
effect of gravitational back-reaction of cosmological fluctuations
is physically measurable, we need to express $\Theta$ in terms of
a physical clock field, not just in terms of the background time.
As was mentioned before, we will use $\chi$ as a clock. Hence, to
obtain results with physical meaning, we must evaluate $\Theta$ on
a surface of constant $\chi$ and not constant $t$.

By dropping the gradient terms from the $G^{00}$ Einstein
equations, it can be shown that
\begin{equation} \label{G00}
\frac{H}{\sqrt{1+2\Phi}} \, = \, {1
\over
{\sqrt{3}}}\bigr[V(\varphi,\chi)+\frac{1}{2}\frac{\dot{\varphi}^{2}}{g_{oo}}+\frac{1}{2}\frac{\dot{\chi}^{2}}{g_{oo}}\bigr]^{1/2}
\,
\end{equation}
where $(1+2\Phi)$ is simply $g_{oo}$, the time-time component of
metric. If we expand the left hand side in terms of $\Phi$ we get
the result,
\begin{equation}
H(1-\Phi+\frac{3}{2}\Phi^{2}) \, = \, {1 \over
{\sqrt{3}}}\bigr[V(\varphi,\chi)+\frac{1}{2}\frac{\dot{\varphi}^{2}}{g_{oo}}+\frac{1}{2}\frac{\dot{\chi}^{2}}{g_{oo}}\bigr]^{1/2}
\, .
\end{equation}
Substituting this result into equation (\ref{expform}) and
neglecting terms containing $\dot{\Phi}$ (which are sub-dominant
compared to other terms in the infrared limit as long as
the equation of state of the background is not changing significantly
over time) leads to Friedmann-like
equation for the local expansion rate
\begin{equation} \label{expform2}
\Theta \, = \, \sqrt{3}
\bigr[V(\varphi,\chi)+\frac{1}{2}\frac{\dot{\varphi}^{2}}{g_{oo}}+\frac{1}{2}\frac{\dot{\chi}^{2}}{g_{oo}}\bigr]^{1/2} \, .
\end{equation}
At this point, it appears at first glance that the expansion rate
for a fixed value of the clock field manifestly depends on the
value of the gravitational potential $\Phi$ which enters via
$g_{oo}$, and that thus the leading infrared back-reaction effects
are locally measurable. However, we are not yet ready to reach
this conclusion since the right hand side of (\ref{expform2})
still contains derivatives with respect to the background
coordinate time which should be re-expressed in terms of local
physical observables.

To describe the behavior of $\Theta$ on surfaces of constant
$\chi$, we have to express the right hand side of (\ref{expform2})
in terms of $\chi$ and of the initial values $\varphi_{in}$ and
$\chi_{in}$ of the matter fields.
This involves solving the Klein-Gordon equations for both scalar fields:
\begin{eqnarray}
V_{,\varphi}(\varphi,\chi)&+&3{H \over
g_{oo}}\dot{\varphi}+\frac{1}{g_{oo}}\ddot{\varphi}=0 \label{klein1}\\
 V_{,\chi}(\varphi,\chi)&+&3{H
\over g_{oo}}\dot{\chi}+\frac{1}{g_{oo}}\ddot{\chi}=0
\label{klein2}\, .
\end{eqnarray}
Since finding exact and analytical solutions of this coupled
system of equations is not possible for a general form of
the potential, we
solved these equations for an infinitesimal period of time $\delta t$.
In other words, we used solutions linear in time to investigate the
dynamic behavior of $\Theta$. We start with initial values of
$\varphi_{in}(x),\dot{\varphi}_{in}, \dot{\chi}_{in}$ at $ t_{in}$
on a surface with a constant value of $\chi_{in}$, let the system
evolve, and calculate the final value of these parameters on a
surface with constant value of $\chi_{f}$. The infinitesimal
evolution equations are
\begin{eqnarray}
\delta t&=&{\chi_{f}-\chi_{in} \over \dot{\chi}_{in}} \\
\dot{\varphi}_{f}&=&\ddot{\varphi}_{in}\delta t+\dot{\varphi}_{in}\\
\varphi_{f}&=&\dot{\varphi}_{in}\delta\eta+\varphi_{in}\\
\dot{\chi}_{f}&=&\ddot{\chi}_{in}\delta t+\dot{\chi}_{in} \, .
\end{eqnarray}
Notice that we do not assign a uniform time to all points on the
initial hypersurface, and that, for now, we just solve equations
locally. Inserting the results from the above equations into the
general expression (\ref{expform2}) for $\Theta_{f}$ leads to
\begin{equation} \label{localtheta}
\Theta_{f}=\Theta_{in}-{3(\chi_f-\chi_{in})\over
2\dot{\chi}_{in}\sqrt{g_{oo}}}(\dot{\varphi}^2_{in}+\dot{\chi}^2_{in})
\, ,
\end{equation}
where the square root of $g_{oo}$ is evaluated on the initial
hypersurface. We are still not finished, because the second term
on the right hand side still contains derivatives with respect to
the coordinate time. So we rewrite the equation (\ref{localtheta})
using the proper time instead of the coordinate time
($\partial\tau=\sqrt{g_{00}}\partial t$), for gauge invariance
purposes,
\begin{equation}
\Theta_f=\Theta_{in}-{3\over2}\chi^{\prime}(1+\frac{\varphi_{in}^{\prime2}}{\chi_{in}^{\prime2}})\delta\chi
\end{equation}
where $\delta\chi = \chi_{f}-\chi_{in}$. Equation
(\ref{localtheta}) can be further simplified using Equations
(\ref{G00}) and (\ref{expform2}), and this enables us to express
$\Theta_{f}$ as a function of $\Theta_{in}$, $\varphi_{in}$,
$\chi$ and $\chi^\prime$,
\begin{equation} \label{localtheta2}
\Theta_{f}=\Theta_{in} - {3 \over 2}(\frac{\Theta^{2}_{in}}{3}
-V(\varphi_{in},\chi_{in}))\frac{\delta\chi}{\chi_{in}^{\prime}}
\, .
\end{equation}
This is the completely local result for the change in $\Theta$
which we were aiming to obtain. One may think that
(\ref{localtheta2}) implies there is no local back-reaction
\footnote{The completely local form of Equation
(\ref{localtheta2}) is easy to understand from the mathematical
point of view: by neglecting all of the spatial derivative terms,
not only are we treating our equations completely locally, but we are also
eliminating all interactions between neighboring points in space.
Thus, the final result for $\Theta$ at a specific point in space
can depend on nothing more than on the initial values of the fields
$\varphi_{in}$ and $\Theta_{in}$ at that same point. In other
words, the dynamics of the fields at each point of space behaves
like a Friedmann Universe with the appropriate initial
conditions.}. However, this conclusion is not correct because in
the presence of cosmological fluctuations the trajectories of the
$\chi$ and $\varphi$ fields start to deviate from the
classical solution that are obtained from Equations (\ref{klein1})
and (\ref{klein2}). Thus, the values of $\varphi_{in}^{\prime}$ and
$\chi_{in}^{\prime}$ for the next time steps will fluctuate on the
$\chi_{in} = {\rm const}$ hypersurface, whereas in the absence of
such fluctuations, the initial values at each step follow the
classical trajectory. In other words, compared to the classical
solution, there is a difference in the evolution of $\Theta$ due
to the different paths for each point of an inhomogeneous universe
in field space. Furthermore, there are other terms coming from
the modes that are exiting the Hubble radius and are not taken in
to account in the infrared limit of Equation (\ref{localtheta2}).
Interpreted correctly, our result (\ref{localtheta2}) thus shows
that one expects that in a two matter field model, infrared
cosmological fluctuations have a locally observable effect if
cosmological perturbations are excited.

To clarify this point we investigate the behavior of
back-reaction terms in two scenarios, first where adiabatic modes
are dominant and then, in the next section, in a scenario where
entropy modes play a major role. In the former case we assume
that the $\chi$ field does not significantly contribute to the energy
density of the universe and only plays the role of a clock. We can
consider $\chi$ to be a rapidly oscillating scalar filed (to mimic
the cosmic microwave background radiation in our Universe) and take
the local energy density of $\chi$ as our clock. For that purpose
we work with the following potential
\begin{equation}
V(\chi)={1\over2}\mu^2 \chi^2 \, ,
\end{equation}
where, by taking  $\mu \gg \Theta$, the rate of oscillation of
$\chi$ will be much faster than the expansion rate, which enables
us to solve Equation (\ref{klein2}) using the WKB approximation
and to calculate the value of $\rho_{\chi} = \chi^{\prime
2}/2+V(\chi)$, the energy density of $\chi$,
\begin{equation}
\rho_{\chi}=\rho_{\chi, in}\exp(-\int_{\tau_{in}}^{\tau}\Theta d\tau)
\, .
\end{equation}
For the inflaton field, we also take a standard quadratic
potential of chaotic inflation,
\begin{equation}
V(\varphi)={1\over2}m^2\varphi^2 .
\end{equation}
To take into account the effects of
the ultraviolet modes that become infrared as
the inflationary comoving Hubble radius shrinks, we modify
Equation (\ref{klein1}) in the slow-roll regime to obtain
\begin{equation} \label{adiabtheta}
\Theta \varphi^{\prime}+m^2 \varphi = C_a \Theta^{5/2} \xi({\bf
x},\tau) \, ,
\end{equation}
where $C_a$ ($a$ standing for adiabatic) is a numerical
constant and $\xi$ is a Gaussian random function with the
covariance
\begin{equation}
\langle\xi({\bf x}_1,\tau_1)\xi({\bf
x}_2,\tau_2)\rangle=\delta(\tau_1-\tau_2){\rm sinc}(aH|{\bf
x}_1-{\bf x}_2|) \, .
\end{equation}

In order to derive Equation (\ref{adiabtheta}) rigorously, one
needs to coarse-grain the non-linear Einstein equations on the
scale of the Hubble radius \footnote{This is similar to the
approach taken in deriving the equations of the stochastic scenario of
inflation \cite{stochast}. A more rigorous derivation which
includes metric perturbations is given in \cite{niayesh}.}.
The stochastic term on the right hand side of Equation \ref{adiabtheta} is
due to the adiabatic fluctuation modes that are exiting the Hubble radius
(see \cite{AG} for a detailed derivation).

Now, using the fact that Equation ({\ref{expform2}) simplifies to
$\Theta=\sqrt{3 V(\varphi)}$ in the slow-roll approximation, one
can solve Equation (\ref{adiabtheta}) perturbatively in $\xi$,
(see \cite{AG} for details) and calculate the expectation value of
$\Theta$,
\begin{equation} \label{thetatau1}
\Theta \simeq \langle\Theta\rangle=m^2(-\tau)+{C_a^2\over
3\sqrt{6}}(\Theta_{in}^3-m^6(-\tau)^3),
\end{equation}
where we have taken $\tau$ to be negative during and to go to zero
at the end of the slow-roll period. This
result shows that back-reaction increases the expansion rate
when proper time is used as the clock. As we shall see, this result
changes when we use $\rho_{\chi}$ as a clock. To simplify
the expression for the expansion rate when $\rho_{\chi}$ is used as
a clock, we define a
new parameter, $\beta=\ln{\rho_{\chi}\over\rho_{fc}}$, which is a
measure of $\rho_{\chi}$. Here, $\rho_{_{fc}}$ is a
constant and the subscript $fc$ refers to the time $\tau_{_{fc}}=0$, the
end of the classic slow-roll inflation. Hence, we have
\begin{equation}
\beta(\tau)=-\int^{\tau}_0\Theta(\tilde{\tau}) d\tilde{\tau}.
\end{equation}
Now, substituting $\Theta$ from Equation (\ref{thetatau1}), in the
above integral we can calculate $\tau$ in terms of $\beta$,
\begin{equation}
-\tau\simeq{\sqrt{2\beta}\over m}\bigl \{1-{C_a^2\over
2\beta\sqrt{6}}({1\over
3}m^4\beta^2+{{\Theta_{in}^3\sqrt{2\beta}}\over 3m}\bigr )\}.
\end{equation}
Substituting $\tau$ back in Equation (\ref{thetatau1}), we can
drive $\Theta$ in terms of $\beta$:
\begin{equation}
\Theta\simeq m\sqrt{2\beta}-{13C_a^2\over
24\sqrt{3}}m^3(\beta)^{3/2}
\end{equation}
Interestingly, the sign of the back-reaction term is negative,
whereas it was positive in Equation (\ref{thetatau1}). Thus, it
appears that when we measure local expansion rate using an
oscillating field as our clock, unlike when we used proper time,
back-reaction slows down the expansion rate. Describing  the
back-reaction in terms of $\beta$ is of more relevance
to the inflationary predictions for today's universe, as
$\beta$ is proportional to number of e-foldings of inflation which
enters into the calculation of the power spectrum of metric fluctuations.

\section{Back-reaction of entropy modes}

In this section we will consider a two matter field model motivated
by hybrid inflation \cite{hybrid}, where it is expected that
the entropy fluctuations are large.
In this section, due to the specific choice of the model that we
investigate, we calculate the back-reaction effect in terms of
proper time, i.e. constant-$\tau$ surfaces, as a clock instead of
constant-$\chi$ surfaces. This is an appropriate choice because we
only want to investigate the magnitude of this effect.
In terms of this clock variable, Equation (\ref{localtheta2})
takes the following form,
\begin{eqnarray} \label{localtheta3}
\Theta_{f}&=&\Theta_{in}-{3\over2}(\chi_{in}^{\prime2}+\varphi_{in}^{\prime2})\delta\tau \nonumber\\
&=&\Theta_{in}
- {3 \over 2}(\frac{\Theta^{2}_{in}}{3}
-V(\varphi_{in},\chi_{in}))\delta\tau \,
\end{eqnarray}
where we see that different trajectories in the $(\varphi,\chi)$
space, which could belong to different points in an inhomogeneous
universe, lead to different behaviors for $\Theta$. As we
demonstrate below, this effect could be clearly seen when we
consider the evolution of $\Theta$ in the long run.

A simple model where we get growing entropy modes is
hybrid inflation, where the secondary field, $\chi$,
has a negative square mass. In the slow-roll regime, the classical
value of $\chi$ vanishes, which is why we do not use it as our
clock. Now, notice that the local evolution of $\Theta$ is given
by
\begin{equation}\label{thetaf}
\Theta_{f} =
\Theta_{in}-{3\over2}\int_{\tau_i}^{\tau_f}(\chi^{\prime2}+\varphi^{\prime2})d\tau.
\end{equation}
If we can show that, due to the presence of entropy modes,
$\chi^{\prime2}$ may dominate $\varphi^{\prime2}$ in Equation
(\ref{thetaf}), we have demonstrated that back-reaction can
dominate the evolution of $\Theta$.

For the sake of clarity, let us investigate the behavior of
$\chi^{\prime2}$ in a model where,
\begin{equation}
V(\varphi,\chi)={1\over 2}m^2\varphi^2-{1\over 2} \mu^2 \chi^2.
\end{equation}
If we assume $\chi$ is also slowly rolling, then, similar to what
happens in Equation (\ref{adiabtheta}), the field equation (\ref{klein2})
should be modified to take into account the entropy fluctuation modes, while
we ignore such modifications (arising due to the adiabatic modes) in Equation
(\ref{klein1}). This leaves us with the following field equations
\begin{eqnarray}
\Theta\chi^\prime-\mu^2\chi&=&C_e\Theta^{5/2}\xi({\bf x},\tau), \\
\Theta\varphi^\prime+m^2\varphi&=&0,
\end{eqnarray}
where $C_e$ ($e$ standing for entropy) is a numerical constant.
Again, using $\Theta \simeq \sqrt{3 V(\varphi)}$, we
can calculate $\chi$ up to first order in $\xi$,
\begin{equation}
\chi \simeq C_e
m^{3}(-\tau)^{1-\mu^{2}/m^{2}}\int_{\tau_{in}}^\tau
(-\tilde{\tau})^{3/2+\mu^{2}/ m^2}\xi({\bf
x},\tilde{\tau})d\tilde{\tau},
\end{equation}
which yields
\begin{equation}
\langle\chi^{\prime2}\rangle=f(\tau)+{C_e^2\mu^4m^2\over
4+2{\mu^2\over m^2}}({\tau_{in}\over
\tau})^{2\mu^2/m^2}\tau_{in}^4,
\end{equation}
where $f(\tau)$ is a correction due to the fluctuations of the
Hubble size modes, and its amplitude decreases with time. For
slow-roll conditions to be satisfied, we need to have
$\tau>{1\over m}$. Inserting this limit for the growing mode in
$\langle\chi^{\prime2}\rangle$ and comparing it with homogeneous
value of $\varphi^{\prime2}$, which is the classical rate of
change in $\Theta$, we get
\begin{equation}
{\langle\chi^{\prime2}\rangle\over\varphi^{\prime2}}\simeq
C_e^2{\mu^4\over
4m^4+2\mu^2m^2}({3\over2})^{2+\mu^2/m^2}\varphi_{in}^{4+2\mu^2/m^2},
\end{equation}
at the end of slow-roll inflation. Since $\varphi_{in}>1$, this relation
implies that if $\mu\gtrsim m$, then back-reaction dominates the
inflation before its end (calculated in the absence of
back-reaction), and the homogenous background solution loses its
validity.

This result can also be important for calculating the spectrum of
fluctuation after inflation ends. We have also seen that since
different points have different initial conditions, they will have
different final expansion rates. To obtain a quantitative measure
of the back-reaction of infrared modes, one needs
to take an average of the expansion rate over the many
initial Hubble patches that form our observable universe.

\section{Conclusions}

In this paper, we have studied the leading back-reaction effects
of long wavelength cosmological fluctuations on a local observable
$\Theta$ which measures the local expansion rate of the Universe.
The observable gives the rate at which neighboring comoving
observers separate and coincides with the usual definition of
expansion in the context of the fluid approach to cosmology. We
considered models with two matter fields, an inflaton field
$\varphi$ which dominates the energy-momentum tensor and leads to
slow-roll inflation, and a second matter field $\chi$ which was
used either as a clock or it contributed to production of entropy
modes. We investigated the behavior of back-reaction terms in two
scenarios, first where adiabatic modes are dominant and then, in a
scenario where entropy modes play a major role. In the former case
to obtain locally measurable statements about the presence or
absence of back-reaction effects, We considered $\chi$ to be a
rapidly oscillating scalar filed (to mimic the cosmic microwave
background radiation in our Universe) and took the local energy
density of $\chi$ as our clock. We evaluated $\Theta$ and found
that not only the back-reacton term does not vanish but the sign
of the back-reaction term is negative as well.In the second
scenario, we analyzed the back-reaction of entropy modes in a
specific cosmological model with negative square mass for the
entropy field and found that back-reaction can become significant
and dominate the evolution of $\Theta$.

Our main result is that the leading infrared terms, the terms
which dominate the effects discussed in \cite{ABM} and \cite{aw0},
are non-vanishing and physically measurable if we use the
second scalar field $\chi$ as a clock. The effect can be particularly
large if the entropy mode of the
cosmological fluctuation field is excited. This confirms the
conjectures made in our previous work \cite{GB1} and is in good
agreement with the physical intuition gained from the study of the
parametric amplification of super-Hubble length cosmological
fluctuations during reheating \cite{bv,fb2}.

In this paper, we have thus shown that infrared back-reaction is
``for real'' in two matter field models. However, how important
the back-reaction is remains to be investigated. Results will be
reported in a subsequent publication\cite{GB3}.

\medskip
\centerline{Acknowledgments}
\medskip

We would like to thank Niayesh Afshordi, Alan Guth, Bojan Losic,
Bill Unruh, Alex Vilenkin and Richard Woodard for many useful
insights during the course of this research. This research is
supported in part (at Brown) by the US Department of Energy under
Contract DE-FG0291ER40688, Task A.

\end{document}